\documentclass[12pt, notitlepage,aps,preprintnumbers,superscriptaddress,nofootinbib,aps,prd,floatfix]{revtex4-2}
\usepackage{float}
\usepackage{array}
\usepackage{mathrsfs}
\usepackage{amsfonts}
\usepackage{epsfig}
\usepackage{amsmath}    % need for subequations
\usepackage{amssymb}   % defines \lesssim, etc
\usepackage{graphicx}   % need for figures
\usepackage{verbatim}   % useful for program listings
\usepackage[dvipsnames]{xcolor}
\usepackage{subfigure}  % use for side-by-side figures
\usepackage{url}
\usepackage{hyperref}
\definecolor{nicered}{rgb}{0.5,0.,0.}
\definecolor{nicegreen}{rgb}{0.,0.5,0.}
\definecolor{niceblue}{rgb}{0.,0.,0.5}
\hypersetup{colorlinks,citecolor=nicegreen,linkcolor=nicered,urlcolor=niceblue}
\usepackage{multirow}
\usepackage{cprotect}
\usepackage{dcolumn}% Align table columns on decimal point
\usepackage{bm}% bold math
\usepackage{slashed}
\usepackage{physics}
\usepackage{framed}

\newcommand{\GeV}{\textrm{GeV}}

\newcommand{\calB}{\mathcal{B}}

\newcommand{\fref}[1]{Fig.\,\ref{fig:#1}} 
\newcommand{\eref}[1]{Eq.~\eqref{eq:#1}} 

\newcommand{\sref}[1]{Sec.~\ref{sec:#1}}

\graphicspath{{./Figures/}}

\makeatother
\usepackage{enumitem}
\setlist{nolistsep}

\begin{document}
\setlength{\abovedisplayskip}{5pt}
\setlength{\belowdisplayskip}{5pt}
\title{Global fits and the 95 GeV diphoton excesses in\\ the Supersymmetric Georgi-Machacek Model}
\author{Yingnan Xu}
\email{yingnanx@smu.edu}
\affiliation{Department of Digital Banking, Qilu Bank, Jinan, Shandong 250014, China\looseness=-1}
\affiliation{Zhongtai Securities Institute for Financial Studies, Shandong University, Jinan, Shandong 250014, China\looseness=-1}
\author{Dikai Li}
\email{lidikai@sztu.edu.cn}
\affiliation{College of Engineering Physics, Shenzhen Technology University, Shenzhen, Guangdong 518118, China\looseness=-1}
\author{Roberto Vega}
\email{rvega@smu.edu}
\affiliation{Department of Physics, Southern Methodist University, Dallas, TX 75275, USA}
\author{Roberto Vega-Morales}
\email{rvegamorales@ugr.es}
\affiliation{Centro Andaluz de Fisica de Particulas Elementales (CAFPE), Universidad de Granada (UGR), Campus de Fuente Nueva, E-18002 Granada, Spain}
\author{Keping Xie}
\email{xiekepi1@msu.edu}
\thanks{Corresponding author}
\affiliation{Department of Physics and Astronomy, Michigan State University, East Lansing, MI 48824, USA\looseness=-1}
%\affiliation{Department of Physics, Southern Methodist University, Dallas, TX 75275, USA}
\preprint{MSUHEP-24-014, SMU-PHY-24-02}
\date{\today}

\begin{abstract}
Recently the ATLAS and CMS experiments have reported modest excesses in the diphoton channel at around 95 GeV.~A number of recent studies have examined whether these could be due to an extended electroweak symmetry breaking (EWSB) sector, including the well known Georgi-Machacek (GM) model.~Here we examine whether the excesses can be explained by a light exotic Higgs boson in the \emph{Supersymmetric} GM (SGM) model which has the same scalar spectrum as the conventional GM model, but with a more constrained Higgs potential and the presence of custodial Higgsino fermions.~We perform a global fit of the SGM model including all relevant production and decay channels, some of which have been neglected in previous studies, which severely constrain the parameter space.~We find that the SGM model can fit the data if the LHC diphoton excesses at 95\,GeV are due to the lightest custodial singlet Higgs boson which contributes $(5-7)\%$ to EWSB, but \emph{cannot} accommodate the LEP $b\bar{b}$ excess, in contrast to other recent studies of the GM model.~Since the SGM model has a highly constrained Higgs potential, the rest of the mass spectrum is sharply predicted, allowing for targeted searches at the LHC or future colliders.~We also compare the SGM model with the non-supersymmetric GM model and identify how they can be distinguished at the LHC or future colliders.

\end{abstract}

\maketitle
%\vspace{-10pt}%Makre sure the table of contents on the first page.
\tableofcontents

\section{Introduction}
\label{sec:intro}

The discovery of a new scalar particle at 125 GeV in 2012 at the Large Hadron Collider (LHC)~\cite{Aad:2012tfa,Chatrchyan:2012xdj} marked a great triumph for the Standard Model (SM)~\cite{Weinberg:1967tq,Salam:1968rm} and confirmed the Higgs mechanism~\cite{Englert:1964et,Higgs:1964pj,Guralnik:1964eu} as responsible for electroweak symmetry breaking (EWSB).~While the observed 125\,GeV scalar appears to have SM like properties~\cite{Falkowski:2013dza}, uncertainties in its coupling measurements~\cite{Khachatryan:2014kca,CMS:2014nkk,Khachatryan:2016vau,Sirunyan:2017exp,Sirunyan:2017tqd,Aaboud:2017oem,Blasi:2017xmc} still leaves room for extended Higgs sectors which can contribute non-negligibly to EWSB.~Many searches for new Higgs bosons have been performed at LEP~\cite{OPAL:2002ifx,LEPWorkingGroupforHiggsbosonsearches:2003ing,ALEPH:2006tnd}, Tevatron~\cite{CDF:2012wzv}, and LHC~\cite{CMS:2015ocq,CMS:2017yta,ATLAS:2018xad,CMS:2018rmh,CMS:2018cyk,ATLAS:2022abz,CMS:2022goy,CMS:2024yhz} and one of the primary goals of current and future collider experiments is to search for additional Higgs bosons above and below 125 GeV.~Recently, the CMS collaboration reported a small diphoton excess around 95 GeV based on data from LHC 8 TeV Run 1~\cite{CMS:2017yta} and 13 TeV Run 2~\cite{CMS:2018cyk} with an integrated luminosity of 19.7 and 35.9\,fb$^{-1}$ respectively.~Subsequently, the ATLAS collaboration also reported an excess with a local significance of $1.7\sigma$ in the diphoton channel around the same mass~\cite{ATLAS:2018xad}\,\footnote{CMS also observed an excess in the di-tau final states, with the mass peaked around 100 GeV~\cite{CMS:2022goy}.~However, this result is somewhat in tension with the recent CMS observation of Higgs production in association with a top quark pair or a $Z$ boson, with subsequent decay into tau pairs~\cite{CMS:2022arx} so is not included.~It also appears to disagree with LEP searches for the process $e^+e^-\to Z\phi(\phi\to\tau\tau)$~\cite{LEPWorkingGroupforHiggsbosonsearches:2003ing} which we also do not include in our analysis.}.~It is interesting to consider the possibility that these LHC diphoton excesses might be the first hints of an extended EWSB Higgs sector.

Of course any extended Higgs sector must be carefully constructed in order to satisfy the stringent constraints~\cite{ParticleDataGroup:2024cfk} from electroweak precision data.~In particular, measurements of the $\rho$ parameter imply the tree level relation $\rho_{tree} = 1$.~As is well known, this is automatically satisfied by Higgs sectors respecting the so called `custodial' global $SU(2)_C$ symmetry~\cite{Sikivie:1980hm}.~Furthermore, explaining the 95\,GeV excess with an additional Higgs boson requires enhanced branching ratios into pairs of photons relative to a SM Higgs at the same mass.~This requires either suppressing the $b\bar{b}$ branching ratio via suppressed couplings to SM fermions and/or enhanced effective couplings to photons.~The first case was considered in a number of studies~\cite{Biekotter:2019kde,Choi:2019yrv,Biekotter:2019gtq,Cao:2019ofo,Biekotter:2021ovi,Biekotter:2021qbc,Heinemeyer:2021msz,Biekotter:2022jyr,Biekotter:2022abc,Iguro:2022dok,Aguilar-Saavedra:2023tql,Cao:2023gkc,Chen:2023bqr,Ahriche:2023wkj,Baek:2024cco} of extended Higgs sectors utilizing electroweak doublets which automatically preserve custodial symmetry.~However, an enhanced diphoton branching ratio is difficult to achieve and additional scalar fields are required to suppress the branching ratio into $b\bar{b}$.~Both an enhanced coupling to photons and a suppressed branching ratio into $b\bar{b}$ can be achieved in models with electroweak triplet scalars since they do not couple to fermions and contain doubly charged scalars.~As was shown in the famous Georgi-Machacek (GM) model~\cite{Georgi:1985nv,Chanowitz:1985ug}, one can add to the SM electroweak doublet one complex and one real electroweak triplet scalar in such a way that custodial symmetry of the Higgs sector is preserved.~This model has long been studied and exhibits a rich phenomenology~\cite{Magg:1980ut,Cheng:1980qt,Lazarides:1980nt,Mohapatra:1980yp,Hartling:2014zca,Chiang:2018cgb,Bairi:2022adc,Ghosh:2022bvz,Chiang:2018cgb,Chen:2022zsh}.  

As in the SM, the GM model exhibits quadratic divergences associated with the extra Higgs boson masses.~Furthermore, there is an additional fine-tuning problem due to the explicit breaking of custodial symmetry by hypercharge as well as Yukawa interactions which give rise to quadratically divergent corrections to the $\rho$ parameter at one-loop~\cite{Gunion:1990dt}.~Typically it is envisioned that the GM model Higgs sector arises as pseudo Goldstone bosons contained in the coset of the global symmetry breaking structure of a strongly coupled sector~\cite{Georgi:1985nv,Bellazzini:2014yua}.~However, it was found more recently that the GM model can also arise as the limit of a weakly coupled supersymmetric theory dubbed the supersymmetric custodial triplet model (SCTM)~\cite{Cort:2013foa,Garcia-Pepin:2014yfa,Delgado:2015aha}.~As shown in~\cite{Vega:2017gkk}, the SCTM has a low energy limit that gives rise to the same Higgs boson sector as in the GM model, but with a much more constrained Higgs potential as well as the presence of light fermionic superpartners in the form of custodial higgsinos.~This low energy limit was dubbed the supersymmetric GM (SGM) model.~Since it arises out of a supersymmetric theory, the SGM solves the various fine tuning problems of the GM model associated with quadratic divergences and also inherits all of the other attractive features of the SCTM~\cite{Cort:2013foa,Garcia-Pepin:2014yfa,Delgado:2012sm,Delgado:2015aha,Delgado:2015bwa,Carena:2013ooa,Carena:2014nza,Carena:2015moc,Garcia-Pepin:2016hvs,Vega:2017gkk}.~Note the custodial symmetry in these GM-like models automatically realizes an ‘alignment’ limit~\cite{Carena:2013ooa,Carena:2015moc} allowing for regions of parameter space which impersonate the SM without decoupling.~Thus we can have light electroweak scalars while remaining consistent with current ``SM-like'' data.~This is also because the custodial symmetry ensures that not only are corrections to $\rho =1$ very small, but the custodial Higgs sector gives rise to compressed (nearly degenerate) mass spectra leading to soft decay products which are challenging to detect at colliders.

Here we perform a global fit of the SGM model to all relevant data including measurements of the observed 125 GeV Higgs boson and find the regions of parameter space where it can fit the 95\,GeV diphoton excesses as well as the rest of current data at least as well as the Standard Model.~As part of our analysis we include the Drell-Yan Higgs pair production channel which, as emphasized in~\cite{Delgado:2016arn,Vega:2018ddp}, is not suppressed even in the limit of small Higgs or VEV mixing since it is mediated by gauge interactions and does not depend on EWSB.~Furthermore, for light Higgs bosons in the small mixing angle limit, Drell-Yan Higgs pair production dominates over single Higgs production channels.~Note this production channel is present in any extension of the EWSB sector, but is not present in the SM so it cannot be obtained by a simple rescaling of the SM production cross section.~It was not considered in previous global fits of GM like models and the 95\,GeV excess~\cite{Chiang:2018cgb,Chen:2022zsh,Chen:2023bqr,Mondal:2025tzi} and is often neglected in experimental searches of exotic Higgs bosons.~We also include $t\bar{t}H_3^\pm \to \tau^\pm \nu$ searches in our analysis which severely constrain the electroweak triplet VEV for light custodial triplet masses~\cite{Ghosh:2022wbe} and which were also not included in previous fits to the 95\,GeV excesses~\cite{Chen:2023bqr}.~Here we include, for the first time in a study of GM-like Higgs sectors, both the Drell-Yan pair production channel and $t\bar{t}H_3^\pm \to \tau^\pm \nu$ searches.~As we'll see, including both of these puts severe constraints on the parameter space\,\footnote{Since Drell-Yan pair production is mediated by gauge interactions it is universal for any extended EWSB sector and is present even in the absence of Higgs or VEV mixing.~Furthermore, even the most minimal extension of the SM Higgs sector (which approximately respects custodial symmetry) includes $H_3^\pm$ so the $t\bar{t}H_3^\pm \to \tau^\pm \nu$ process will also be present.~Thus the severe constraints obtained here for the SGM model should also be found in fits of any model of an extended EWSB sector.}.~Since the SGM model has in addition a highly constrained Higgs potential~\cite{Vega:2018ddp}, the rest of the scalar and fermion mass spectrum is sharply predicted allowing for targeted search strategies at colliders.

We find that the SGM model can accommodate the data if the 95\,GeV diphoton excesses are due to the lightest custodial singlet Higgs boson which contributes $(5-7)\%$ to EWSB.~Since the SGM model has a highly constrained Higgs potential, the rest of the mass spectrum is sharply predicted.~In particular, the SGM model predicts a doubly charged scalar in the mass range $\sim(185 - 195)$\,GeV as well as a doubly charged fermion in the range $\sim(175 - 220)$\,GeV where both are associated with a custodial fiveplet~\cite{Vega:2017gkk}.~The global fit also points to a fermion LSP, which is the neutral component of the custodial triplet Higgsino, with a mass in the range $\sim(117 - 137)$\,GeV.~We briefly comment on potential signatures and search strategies for uncovering the spectrum of the SGM model at a high luminosity LHC or future high energy collider.~We also compare the parameter space in the SGM model with a constrained version of the non-supersymmetric GM model where we constrain the GM model Higgs potential in the same way as the SGM model in order to isolate the effects of the custodial higgsino fermions on the custodial Higgs boson phenomenology.

%%%%%%%
%%%%%%%
%%%%%%%
\section{Review of the Supersymmetric Georgi-Machacek model}
\label{sec:model}

Here we briefly review the relevent features for our current analysis of the Supersymmetric Georgi-Machacek (SGM) model while further details can be found in~\cite{Cort:2013foa,Vega:2017gkk}.~The SGM model contains the same Higgs sector as the well known Georgi-Machacek (GM) model~\cite{Georgi:1985nv,Chanowitz:1985ug} with scalar multiplets of custodial symmetry, but with additional custodial (higgsino) fermions at around the same scale.~At the weak scale the SGM is not a supersymmetric theory since it contains only half of the necessary scalar degrees of freedom in the Higgs sector, but it arises as a limit~\cite{Vega:2017gkk} of the fully supersymmetric theory dubbed the Supersymmetric Custodial Triplet Model (SCTM)~\cite{Cort:2013foa,Garcia-Pepin:2014yfa,Delgado:2015bwa},\,constructed to alleviate the MSSM Higgs mass `problem' while also satisfying constraints from electroweak precision data (EWPD) and other direct searches.~Thus the SGM model is free of additional fine tuning problems associated with extra scalars and that are inherent to the conventional GM model~\cite{Gunion:1990dt}.

%%%%%%%
%%%%%%%
%%%%%%%
\subsection{Relating the SGM to the GM model}
\label{sec:mapping}

In addition to the SM complex scalar electroweak doublet $\phi=(\phi^{+},\phi^{0})$ with hypercharge $Y=1$\footnote{We take the convention that $Q=T_3+Y/2$.}, the Georgi-Machacek model~\cite{Georgi:1985nv,Chanowitz:1985ug} also consists of one complex scalar triplet $\chi=(\chi^{++},\chi^{+},\chi^{0})$ with hypercharge $Y=2$ and one real scalar triplet $\xi=(\xi^{+},\xi^{0},\xi^{-})$ with hypercharge $Y=0$.  These fields can be combined to form the $(2_L,2_R)$ and $(3_L,3_R)$ representations~\cite{Low:2010jp} under the global $SU(2)_L\otimes SU(2)_R$ symmetry, respectively,
\begin{equation}\label{eq:GMscalars}
\Phi=\begin{pmatrix}
\phi^{0*} & \phi^{+}\\
\phi^{-} & \phi^{0}
\end{pmatrix},~
\Delta=\begin{pmatrix}
\chi^{0*} & \xi^{+} & \chi^{++}\\
\chi^{-} & \xi^{0} & \chi^{+}\\
\chi^{--} & \xi^{-} & \chi^{0}
\end{pmatrix},
\end{equation}
which transform under $SU(2)_L\otimes SU(2)_R$ as,
\begin{equation}\label{eq:transf}
\Phi\to U_{L}\Phi U_{R}^{\dagger}, ~ \Delta \to U_{L}\Delta U_{R}^{\dagger}.
\end{equation}
The most general scalar Lagarangian under the SM $SU(2)_L\otimes U(1)_Y$ gauge symmetry as well as the $SU(2)_L\otimes SU(2)_R$ global symmetry can then be written as,
\begin{equation}
\mathcal{L}=\frac{1}{2}\Tr[(D^{\mu}\Phi)^{\dagger}D_{\mu}\Phi]+\frac{1}{2}\Tr[(D^{\mu}\Delta)^{\dagger}D_{\mu}\Delta]-V(\Phi,\Delta),
\end{equation}
where $D_{\mu}$ is the covariant derivative.~The Higgs potential is given by~\cite{Hartling:2014zca},
\begin{equation}\label{eq:VGM}
\begin{aligned}
V(\Phi,\Delta)
&=\frac{1}{2}m_{\phi}^{2}\Tr[\Phi^{\dagger}\Phi]+\frac{1}{2}m_{\Delta}^{2}\Tr[\Delta^{\dagger}\Delta]
+\lambda_{1}\left(\Tr[\Phi^{\dagger}\Phi]\right)^{2}+\lambda_{2}\Tr[\Phi^{\dagger}\Phi]\Tr[\Delta^{\dagger}\Delta]\\
&+\lambda_{3}\Tr[\Delta^{\dagger}\Delta\Delta^{\dagger}\Delta]+\lambda_{4}\left(\Tr[\Delta^{\dagger}\Delta]\right)^{2}
-\lambda_{5}\Tr[\Phi^{\dagger}\tau^{a}\Phi\tau^{b}]\Tr[\Delta^{\dagger}t^{a}\Delta t^{b}]\\
&-M_{1}\Tr[\Phi^{\dagger}\tau^{a}\Phi\tau^{b}]\left(U\Delta U^{\dagger}\right)_{ab}
-M_{2}\Tr[\Delta^{\dagger}t^{a}\Delta t^{b}]\left(U\Delta U^{\dagger}\right)_{ab},
\end{aligned}
\end{equation}
where $\tau^{a}=\sigma^{a}/2$ (with $\sigma^{a}$ being the Pauli matrices) and $t^{a}$ are the $SU(2)$ generators for the doublet and triplet representations respectively.~The matrix $U$ rotates the bi-triplet $\Delta$ into the Cartesian basis and is given by,
\begin{equation}
U=\frac{1}{\sqrt{2}}\begin{pmatrix}
-1 & 0 & 1\\
-i & 0 & -i\\
0 & \sqrt{2} & 0
\end{pmatrix}.
\end{equation}
After EWSB, the scalar fields develop non-zero vacuum expectation values (VEVs),
\begin{equation}
\langle\Phi\rangle=\begin{pmatrix}
v_{\phi} & 0\\
0 & v_{\phi}
\end{pmatrix},~
\langle \Delta\rangle=\begin{pmatrix}
v_{\chi} & 0 & 0\\
0  & v_{\xi} & 0\\
0 & 0 & v_{\chi}
\end{pmatrix} .
\end{equation}
When $v_\xi=v_\chi\equiv v_\Delta $,~i.e.~the triplet VEVs are aligned, then the $SU(2)_L\otimes SU(2)_R$ symmetry will be broken to the custodial $SU(2)_C$ diagonal subgroup ensuring that $\rho_{tree} = 1$ as in the SM~\cite{Low:2010jp}.~The bi-doublet and bi-triplet Higgs fields then decompose under the custodial $SU(2)_C$ as $({\bf2,\bar{2}}) = {\bf 1\oplus 3}$ and $({\bf3,\bar{3}}) = {\bf 1\oplus 3 \oplus 5}$.~After EWSB the two custodial singlets can mix as can the two custodial triplets one of which becomes the three Goldstone bosons eaten by the electroweak gauge bosons.~The electroweak doublet and triplet VEVs are then related to the $W$ boson mass and electroweak scale as, 
\begin{equation}\label{eq:VEVs}
v_{\phi}^{2}+8v_{\Delta}^{2}\equiv v^{2}=\frac{4M_W^2}{g_2^2}=\frac{1}{\sqrt{2}G_F}\simeq(246~\GeV)^{2},
\end{equation}
where $g_2$ is the SM $SU(2)$ gauge coupling and $G_F$ is the Fermi constant.~This global symmetry breaking structure can also be imbedded into composite Higgs models~\cite{Georgi:1985nv,Chanowitz:1985ug,Bellazzini:2014yua}.~Using the relation in~\eref{VEVs},\,we can parameterize the electroweak doublet and triplet VEVs in terms of a VEV mixing angle ($\theta_H$) and electroweak scale ($v$) as,
\begin{equation}\label{eq:VEVmixing}
c_H = \frac{v_\phi}{v}, ~ s_H = \frac{2\sqrt{2}v_\Delta}{v},
\end{equation}
where $s_H \equiv \sin\theta_H,c_H \equiv \cos\theta_H$ and $s_H = 0$ corresponds to zero electroweak triplet VEV. 

As shown in~\cite{Vega:2017gkk}, the Higgs potential in~\eref{VGM} can be `derived' as a limit from the Higgs superpotential and SUSY breaking sector of the $SU(2)_L\otimes SU(2)_R$ symmetric  SCTM~\cite{Cort:2013foa,Garcia-Pepin:2014yfa,Delgado:2015bwa}.~The supersymmetric origins of the SGM model then impose a relation between the quartic couplings~\cite{Vega:2017gkk} in the Higgs potential,
\begin{equation}
\begin{aligned}
\label{eq:gmcon}
\lambda_1 = \frac{3}{4} \lambda_2,\quad 
\lambda_3 = -\lambda_4, \quad
\lambda_5 = -4\lambda_2 + 2\sqrt{2\lambda_2\lambda_4} ,
\end{aligned}
\end{equation}
which reduces the number of quartic couplings from five to two.~Furthermore, holomorphy of the SCTM superpotential implies~\cite{Vega:2017gkk}, 
\begin{equation}
\begin{aligned}\label{eq:holo1}
\lambda_2 > 0,\quad
\lambda_4 > 0 .
\end{aligned}
\end{equation}
Together with the constraint on the Higgs quartic couplings in~\eref{gmcon}, this leads to
\begin{equation}
\begin{aligned}\label{eq:holo2}
\lambda_1> 0,\quad  \lambda_3 < 0,\quad
\lambda_5\begin{cases}  
<  0,~\textrm{if}~\lambda_4 < 2  \lambda_2, \\
>  0,~\textrm{if}~\lambda_4 > 2  \lambda_2 . 
\end{cases}
\end{aligned}
\end{equation}
Note however, in the conventional GM model without a supersymmetric origin, $\lambda_2$ and $\lambda_4$ can (both) be negative even if the constraint in~\eref{gmcon} holds by some coincidence.~This implies the signs of the quartic couplings may perhaps be a way to distinguish between a supersymmetric and non-supersymmetric origin for the GM model.

Following this discussion, we see the SGM model can be defined as a weak scale effective theory given by the same custodial Higgs sector as the GM model, but with the constraints in~\eref{gmcon} and~\eref{holo1} applied (which then implies~\eref{holo2}), plus custodial fermions at around the same scale with masses determined by the parameters of the Higgs potential.~We also define what we call the `constrained' GM (CGM) model which does not contain a fermion sector and in which we impose~\eref{gmcon} - \eref{holo2} on the GM potential by hand.~In the end we are left with six independent parameters in the Higgs potential,
\begin{equation}\label{eq:SGM2GM}
(\lambda_2,\lambda_4,M_1,M_2,v_\phi,v_\Delta),
\end{equation}
where we have traded in $m_\phi^2$ and $m_\Delta^2$ for $v_\phi$ and $v_\Delta$ in the Higgs potential (see~\eref{VGM}) by imposing the vacuum minimization conditions~\cite{Cort:2013foa}.~With these parameters as input the physical mass spectrum of both the scalars and fermions along with their couplings can be determined in the SGM and CGM models.~Since these two models have the same Higgs potential, the phenomenology of the custodial Higgs bosons is very similar and the only way to distinguish them is through the effects of the fermion sector in the SGM model.

%%%%%
%%%%%
\subsection{The Higgs scalar mass spectrum}
\label{sec:sspec}

In the custodial basis after EWSB, the Higgs scalars in the SGM and GM models can be decomposed into one fiveplet $H_5^{0,\pm,\pm\pm}$, two triplets $H_3^{0,\pm},G^{0,\pm}$, and two singlets $H_1^0,H_1^{0'}$ of the custodial $SU(2)_c$.~Here $G^{0,\pm}$ are the massless Goldstone bosons eaten by the gauge bosons $Z,W^{\pm}$ to acquire mass.~The squared masses of the fiveplet and triplet are given by,
\begin{equation}\label{eq:m35}
\begin{aligned}
m_5^2& = \frac{M_1}{4v_\Delta}v_\phi^2
+ 12 M_2v_\Delta+\frac{3}{2}
(-4\lambda_2 + \sqrt{2\lambda_2\lambda_4}) 
v_\phi^2 - 8\lambda_4 v_\Delta^2,\\
m_3^2&=
\left(\frac{M_1}{4v_\Delta}
- 2\lambda_2 + \frac{1}{2} \sqrt{2\lambda_2\lambda_4} 
\right)(v_\phi^2 + 8v_\Delta^2)
= \left(\frac{M_1}{4v_\Delta} 
- 2\lambda_2 + \frac{1}{2} \sqrt{2\lambda_2\lambda_4}
\right)v^2,
\end{aligned}
\end{equation}
where we have imposed the supersymmetric constraints in~\eref{gmcon} - \eref{holo2} on the Higgs potential parameters.~The custodial singlets $H_1^0,H_1^{0'}$ mix together leading to the mass matrix,
\begin{equation}
\mathcal{M}^2=\begin{pmatrix}
A & B \\
B & C
\end{pmatrix},
\end{equation}
where we have defined,
\begin{equation}
\begin{aligned}
A&=6 \lambda_{2} v_{\phi}^{2},\quad 
B= -\frac{\sqrt{3}}{2} v_{\phi}\left(M_{1}+8 (2 \lambda_{2} + \sqrt{2 \lambda_2\lambda_4}) v_{\Delta}\right),\\
C&=\frac{M_{1} v_{\phi}^{2}}{4 v_{\Delta}}-6 M_{2} v_{\Delta}+6 \lambda_{4} v_{\Delta}^{2}.
\end{aligned}
\end{equation}
Rotating by a custodial singlet Higgs mixing angle defined as,
\begin{equation}
\begin{aligned}\label{eq:hHmixing}
\sin 2 \alpha=\frac{2 B}{m_{h}^{2} - m_{H}^{2}},\quad  
\cos 2 \alpha=\frac{(C-A)}{m_{h}^{2} - m_{H}^{2}} , \quad
\alpha \in \left(-\frac{\pi}{2},\,\frac{\pi}{2}\right) ,
\end{aligned}
\end{equation}
we can diagonalize the mass matrix and obtain the physical masses as,
\begin{equation}\label{eq:Higgsmix}
m_{h, H}^{2}=\frac{1}{2}\left[A+C \pm \sqrt{(A-C)^{2}+4 B^{2}}\right].
\end{equation}
Considering the discovery of a Higgs boson at 125 GeV~\cite{Aad:2012tfa,Chatrchyan:2012xdj} with SM like properties~\cite{ATLAS:2022vkf,CMS:2022dwd}, one of these custodial singlets should be interpreted as the SM-like Higgs boson implying it is composed mostly of the electroweak doublet while the other custodial singlet will be mostly electroweak triplet.~In this work we will explore the possibility that the $m_h=125$\,GeV Higgs boson is the heavier of the custodial singlets while the lighter singlet we will associate with a possible light scalar resonance at $m_H=95$\,GeV.

%%%%%
%%%%%
%%%%%
\subsection{The higgsino fermion mass spectrum}
\label{sec:fspec}

In the SGM model~\cite{Vega:2017gkk} there is the presence of a neutralino/chargino fermion sector reflecting its supersymmetric origins~\cite{Cort:2013foa,Garcia-Pepin:2014yfa,Delgado:2015bwa}.~Like the scalar Higgs bosons, these fermions (mostly Higgsino) can be arranged into a custodial singlet and triplet coming from the (MSSM) electroweak doublets and a custodial singlet, triplet, and fiveplet coming from the electroweak triplet superfields.~As in the Higgs sector, after EWSB the two custodial singlets can mix as can the two triplets.~The Higgsino masses are determined by the Higgs potential parameters in~\eref{VGM} and are therefore correlated with the Higgs scalar masses.~This is important because it implies the custodial higgsinos cannot be decoupled without decoupling the custodial Higgs sector as well.~Note this also implies that one cannot recover the CGM model as a decoupling limit of the SGM.~Thus, if the 95\,GeV excess is due to the lighter custodial singlet Higgs boson in the SGM model, there is an upper bound on the higgsino masses not too far above the weak scale, as we'll see below.~There are also the gauginos with independent masses which we take to be much heavier than the weak scale.

In the SGM model~\cite{Vega:2017gkk}, the Higgsino masses depend on the electroweak doublet and triplet $\mu$ and $\mu_\Delta$ terms (as well as EWSB) in the SCTM superpotential respectively.~Since the Higgs potential in~\eref{VGM} is derived from the superpotential (and SUSY breaking sector), we can relate these $\mu$ terms to the trilinear mass terms of the GM Higgs potential as~\cite{Vega:2017gkk},
\begin{equation}
\begin{aligned}\label{eq:muterms}
\mu = - \Big(  \frac{M_1}{8\sqrt{\lambda_2}} 
+ \frac{M_2}{2\sqrt{2\lambda_4}}  \Big), \quad
\mu_\Delta = -  \frac{M_2}{\sqrt{2\lambda_4}} .
\end{aligned}
\end{equation}
We can then write the mass of the custodial fiveplet fermion as,
\begin{equation}\label{eq:Mf5}
M_{f_5} = \sqrt{\lambda_4}\,v_{\Delta}
+ \mu_\Delta .
\end{equation}
Since the electroweak triplet VEV must be small to be consistent with measurements of the SM-like 125\,GeV Higgs boson, the fiveplet higgsino mass gives a direct measure of the $\mu_\Delta$ term in the SCTM superpotential.~The singlet neutralino (higgsino) mass matrix becomes\,\footnote{For universal gaugino masses $M_{\tilde{B}}=M_{\tilde{W}}=M$, the photino $\tilde{\gamma}$ does not mix with the custodial singlet higgsinos (or zino) and has a mass $M_{\tilde{\gamma}}=M$~\cite{Vega:2017gkk,Xie:2019eoe}.},
\begin{equation}\label{eq:Mf1}
M_{f_1}=\begin{pmatrix}
3\sqrt{\lambda_2/2}\,v_{\Delta}
 - \mu & 
\sqrt{3\lambda_2}\,v_\phi \\
\sqrt{3\lambda_2}\,v_\phi &
\mu_\Delta 
- 2 \sqrt{\lambda_4}\,v_{\Delta}
\end{pmatrix}.
\end{equation}
The two custodial triplet higgsinos also mix with the wino analagously to how the custodial Higgs triplet mixes with the Goldstone bosons eaten by the $W$ and $Z$ bosons.~This leads to,
\begin{equation}\label{eq:Mf3}
M_{f_3}=
\begin{pmatrix}
M & 
G\,v_\phi/\sqrt{2} & 
\sqrt{2}G v_{\Delta} \\
G\,v_\phi/\sqrt{2}& 
\sqrt{\lambda_2/2}\,v_{\Delta} +\mu & 
-\sqrt{2\lambda_2} v_\phi \\
\sqrt{2}G v_{\Delta} & 
-\sqrt{2 \lambda_2}  v_\phi & 
\sqrt{\lambda_4}\,v_{\Delta} -\mu_{\Delta}
\end{pmatrix},
\end{equation}
where $G=\sqrt{g'^2+g^2}$ for the neutral components and $G=g$ for charged ones respectively.~Thus, there is a small mass splitting between the neutral and charged components of the custodial triplet fermion reflecting the explicit breaking of custodial symmetry due to the hypercharge gauge interactions.~Note in the scalar sector this is reflected in the difference between the $W^{\pm}$ and $Z$ masses whose longitudinal components are the custodial triplet Goldstone bosons as discussed above.~In general these fermions can be produced in pairs via Drell-Yan, but can be difficult to detect because of their compressed spectra~\cite{Schwaller:2013baa,Ismail:2016zby} leading to soft decay products.~Thus they are only constrained by direct searches to be $\gtrsim 100$~GeV and perhaps even as low as $\sim 75$~GeV~\cite{Egana-Ugrinovic:2018roi}.~We take the gaugino mass $M$ to be much larger than the weak scale so we are left with mostly Higgsino custodial triplets at the weak scale who's neutral component will constitute the lightest stable particle (LSP).~Over some regions of parameter space, the LSP can make a viable thermal dark matter candidate~\cite{Delgado:2015aha}.

%%%%%%%%%%%
%%%%%%%%%%%
%%%%%%%%%%
\section{Model constraints}
\label{sec:constraints}

In this section, we will review the various theoretical and experimental (both direct and indirect) constraints on the CGM and SGM models.~This includes the 95 GeV excesses as well as LHC measurements of the 125 GeV Higgs boson couplings.~Our procedure for accounting for these bounds in our parameter scan follows closely the one presented in~\cite{Chen:2023bqr} to which we refer the read for more details. 

\subsection{Theoretical constraints}

To ensure consistency of the CGM and SGM models, there are several theoretical constraints which the parameters of the Higgs potential in~\eref{VGM} must satisfy. 

\textbf{Supersymmetry:} As detailed in~\cite{Vega:2017gkk}, supersymmetry and holomorphy of the superpotential implies the constraints on the quartic couplings given in~\eref{gmcon} -~\eref{holo2} and which fixes their signs.~In the case of the CGM model we impose these constraints by hand. 

\textbf{Perturbative Unitarity:} The perturbative unitarity requires the zeroth partial wave amplitude $a_0$ satisfy $|a_0|\leq1$ or $|\Re(a_0)|\leq1/2$. The $2\to2$ and leads to the bounds on the scalar quartic couplings found in~\cite{Aoki:2007ah,Hartling:2014zca} which we impose here.

\textbf{Boundedness from below (BFB):} The scalar potential should be bounded from the below which imposes additional bounds on the quartic couplings~\cite{Hartling:2014zca}.~For the SGM model this is automatic since it arises from a supersymmetric theory.  

\textbf{Vacuum stability:} Stability of the vacuum requires the custodialy symmetric vacuum to be the unique global minimum of the scalar potential.~This can be achieved through a numerical scanning of the different combinations of the triplet VEVs, $v_\chi$ and $v_\xi$, and ensuring the custodial vacuum (with $v_\xi = v_\chi = v_\Delta$) to be the global minimum~\cite{Hartling:2014zca}.~In addition, tachyonic states~\cite{Chen:2022ocr} corresponding to saddle points of the scalar potential can be avoided by requiring the squared masses of all scalars to be positive.

\subsection{Experimental bounds}
\label{sec:exp}

There are also numerous relevant experimental constraints that the SGM and CGM models must satisfy from both direct and indirect searches at colliders and lower energy experiments as well as measurements of the 125\,GeV Higgs boson at the LHC.~Starting with the 95\,GeV excesses, the CMS collaboration reported a small diphoton excess around 95\,GeV based on the data of LHC 8 TeV Run 1~\cite{CMS:2017yta} and 13 TeV Run 2~\cite{CMS:2018cyk,CMS:2023yay} with an integrated luminosity of 19.7\,fb$^{-1}$ and 132.2\,fb$^{-1}$ respectively.~Subsequently, the ATLAS collaboration also reported an excess with a local significance of $1.7\sigma$ in the di-photon channel around the same mass~\cite{ATLAS:2018xad}.~Neglecting possible correlations, the combined signal strength for this excess in inclusive diphoton searches was obtained~\cite{Biekotter:2023oen},
\begin{equation}
\mu_{\gamma \gamma}^{\rm LHC}=
\frac{\sigma^{\exp }(gg\rightarrow \phi \rightarrow \gamma \gamma)}{\sigma^{\mathrm{SM}}(gg\rightarrow H \rightarrow \gamma \gamma)}
= \mu_{\gamma \gamma}^{\mathrm{ATLAS}+\mathrm{CMS}}=0.24_{-0.08}^{+0.09},
\end{equation}
where $\sigma^{\mathrm{SM}}$ denotes the cross section for a SM-like Higgs boson at the same mass.~This signal strength corresponds to an excess of $3.1\sigma$ at a mass,
\begin{equation}
\begin{aligned}
m_\phi \equiv m_\phi^{\mathrm{ATLAS}+\mathrm{CMS}} 
= 95.4~\mathrm{GeV}.
\end{aligned}
\end{equation}
Previous experiments at LEP also reported an excess in the process, $e^{+} e^{-} \rightarrow Z\phi(\phi \rightarrow bb)$, with a local significance of $2.3\sigma$~\cite{LEPWorkingGroupforHiggsbosonsearches:2003ing}.~The corresponding signal strength at 95 GeV was determined to be~\cite{Cao:2016uwt,Azatov:2012bz},
\begin{equation}
    \mu_{bb}^{\rm LEP}=0.117\pm0.057.
\end{equation} 
As we'll see below, and in contrast to the results found in~\cite{Chen:2023bqr}, due to the stringent constraints on the electroweak triplet VEV and the presence of Drell-Yan pair Higgs production which were not included in~\cite{Chen:2023bqr}, this excess cannot be fit in the SGM and CGM models so in this context the LEP excess is interpreted as a statistical fluctuation. 

To assess the compatibility of the SGM and CGM models with direct searches we can define signal strength for a new scalar resonance as,
\begin{equation}
\begin{aligned}\label{eq:mu}
\mu_i &=
\frac{[\sigma(X_i)\times\calB_i]_{\rm model}}{[\sigma_{\rm ggF} \times \calB]_{\rm SM}}
= (\kappa_{f,i}^2  +  r_{\rm{DY}}+r_{V} )
\frac{\calB_{i,\rm model}}{\calB_{i,\rm SM}},  \\
\kappa_{f,i}^2 &= \frac{\sigma_{\rm ggF}(X_i)}{\sigma_{\rm ggF}} ,\quad
r_{\rm{DY}} \equiv\frac{ \sigma_{\rm{DY}} }{\sigma_{\rm ggF}}\kappa_{VSS}^2 ,\quad
r_{V}\equiv \frac{\sigma_{\rm VBF+VH}}{\sigma_{\rm ggF}}\kappa_{V,i}^{2},
\end{aligned}
\end{equation}
where $\calB_i$ are the branching fractions and $\kappa_i^2$ (the so called $\kappa$ factors~\cite{LHCHiggsCrossSectionWorkingGroup:2012nn}) represents the ratio of cross sections for production channels which also exist in the SM, and $\kappa_{VSS}$ corresponds to the vector-scalar-scalar interaction (largely due to gauge interactions).~Thus we can equate them with the couplings of the scalar to SM particles normalized to the SM couplings.~As discussed above, we also include the Drell-Yan (DY) Higgs pair production channel which is not present in the SM, but can dominate for light enough Higgs bosons.~The theoretical prediction for DY production is calculated at the QCD next-to-the-leading order (NLO) using \texttt{MadGraph\_aMC@NLO}~\cite{Alwall:2014hca,Frederix:2018nkq} interfaced with the GM UFO model files~\cite{Degrande:2015xnm} while the efficiency is extracted form the simulation in~\cite{Ismail:2020zoz,Ismail:2020kqz}.~The cross sections for gluon-gluon fusion (ggF) and vector-boson associated with Higgs production (VH) is calculated up to N3LO with \texttt{n3loxs}~\cite{Baglio:2022wzu} while the vector-boson fusion (VBF) production cross sections are calculated with \texttt{proVBFH}~\cite{Dreyer:2016oyx} at N3LO and the parton distribution functions are taken as CT18NNLO~\cite{Hou:2019efy}.~Replacing the SM prediction $[\sigma\times\calB]_{\rm SM}$ in Eq.~(\ref{eq:mu}) with the experimental exclusion limits, we can test whether a model parameter point is allowed by the corresponding searches.~This methodology is adopted in the \texttt{HiggsBounds}~\cite{Bechtle:2008jh,Bechtle:2011sb,Bechtle:2013wla,Bechtle:2012lvg,Bechtle:2020pkv} sub-package of~\texttt{HiggsTools}~\cite{Bahl:2022igd} which we rely on to perform a global analysis of existing constraints.~The complete list of the measurements included in our analysis can be found in Tabs.\,1-2 of the~\texttt{HiggsTools} manual~\cite{Bahl:2022igd}.

To ensure that the allowed parameter regions are consistent with the rest of LHC data, we employ the \texttt{HiggsTools} analysis package~\cite{Bahl:2022igd} to incorporate the relevant experimental constraints on the extended Higgs sector.~For the measured properties of the observed 125\,GeV Higgs boson we employ the sub-package \texttt{HiggsSignals}~\cite{Bechtle:2013xfa,Bechtle:2020uwn} and to include searches for additional Higgs bosons at the LHC and at LEP we use the \texttt{HiggsBounds} sub-package~\cite{Bechtle:2008jh,Bechtle:2011sb}.~These packages include all the relevant datasets from the LHC Run 2.~However, we have added to the definition of the signal strength the DY Higgs pair production channel, which is not included in the \texttt{HiggsTools} package, as well as the VBF and VH production channels.~We also include the indirect constraint from measurements of $BR(b\to s\gamma)$~\cite{ParticleDataGroup:2024cfk} (as described in \texttt{GMCalc}~\cite{Hartling:2014xma}) as well as $t\bar{t}H_3^\pm \to \tau^\pm \nu$ searches~\cite{CMS:2019bfg}.For the SGM model we must also include the relevant constraints from LEP and LHC searches for heavy lepton or neutralino/chargino searches~\cite{Egana-Ugrinovic:2018roi,ATLAS:2017vat,ATLAS:2019lng,CMS:2019zmn}.

%%%%%%%
%%%%%%%
%%%%%%%
\section{Analysis results}
\label{sec:results}

In this section, we present our results for the global parameter scan to quantify the ability of the SGM model to describe the 95\,GeV excesses and accommodate 125\,GeV Higgs data as well as constraints from direct and indirect searches.~As part of our parameter scans of the SGM, we perform a comparison between the SGM and CGM models in order to study the effects of the fermion superpartners in the SGM model and identify the best way to distinguish between the models.~As we'll see, the primary effect of the fermion superpartners is through loops of charged fermions entering in the $h$ and $H$ coupling to photons. 

Explicit expressions for all of the couplings in the SGM~\cite{Vega:2017gkk} and GM models can be found in~\cite{Hartling:2014zca} and~\cite{Cort:2013foa} respectively.~The full SCTM Lagrangian has been implemented into the~\texttt{SPheno}~\cite{Porod:2003um,Porod:2011nf} and~\texttt{SARAH}~\cite{Staub:2008uz,Staub:2013tta,Staub:2015kfa} public codes verifying the results.~These are used to calculate the full spectrum as well as production cross sections and decay widths, including loop induced decays, of the GM like scalars and their fermionic superpartners in the SGM model.~We have also implemented the GM model into the~\texttt{SPheno}~\cite{Porod:2003um,Porod:2011nf} and~\texttt{GMCalc}~\cite{Hartling:2014xma} codes in order to validate the GM spectrum of the SGM model as well as the corresponding mixing matrices and couplings.

%%%%%%%%%%%
%%%%%%%%%%%
%%%%%%%%%%%
\subsection{Scan over SGM Higgs potential parameters}\label{sec:SGMscan}

\begin{figure}
\centering
\includegraphics[width=0.99\textwidth]{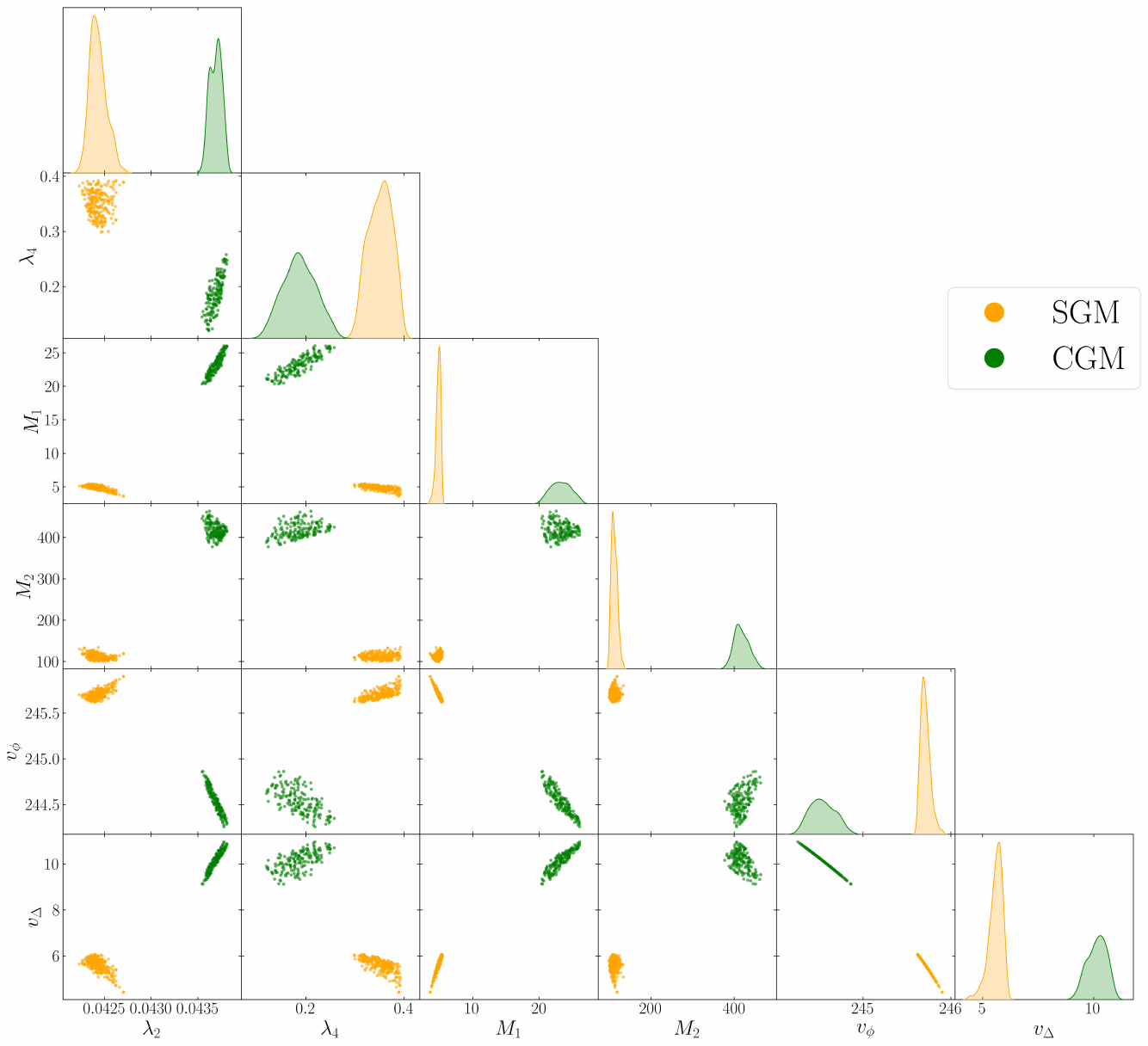}
\caption{Results of the (three dimensional) parameter scan over Higgs potential parameters $(\lambda_2,\,\lambda_4,\,M_1,\,M_2,\,v_\phi,\,v_\Delta)$ for the SGM (yellow) and CGM (green) models.}
\label{fig:params}
\end{figure}
%%%%%
In principle the neutral component of any of the custodial Higgs bosons could be responsible for the 95\,GeV diphoton excesses.~However, the custodial fiveplet and triplet at 95\,GeV are ruled out by a combination of the direct searches for charged and neutral Higgs bosons as well as indirect searches.~Taking the 95\,GeV diphoton excess to come from the lighter custodial singlet Higgs boson, we first fix in our scans $m_H = 95$\,GeV,\,$m_h = 125$\,GeV,\,and $v = 246$\,GeV.~This leaves us with three independent input parameters in the Higgs potential in~\eref{VGM} which then determine the other three.~Imposing all theoretical and experimental constraints discussed above, we show in~\fref{params} the parameter space of the Higgs potential parameters $(\lambda_2,\,\lambda_4,\,M_1,\,M_2,\,v_\phi,\,v_\Delta)$ in the SGM (yellow) and CGM (green) models which fits the LHC diphoton excesses and is consistent with the rest of the data at least as well as the SM (at 68\% C.L.).

For the quartic couplings we see the scan prefers $\lambda_2 \sim 0.0425,\,\lambda_4 \sim (0.3 - 0.4)$ in the SGM model (yellow) and $\lambda_2 \sim 0.0435,\,\lambda_4 \sim (0.1 - 0.3)$ in the GM model (green).~For the mass parameters the scan prefers $M_1 \sim 5\,{\rm GeV},\,M_2 \sim\,100\,\rm{GeV}$ in the SGM model and $M_1 \sim 25\,{\rm GeV},\,M_2 \sim\,400\,{\rm GeV}$ in the CGM model.~So we see while the scan prefers similar quartic couplings in the SGM and CGM models, the preferred mass parameters are $\sim$ (4 - 5) times larger in the CGM model than in the SGM model.~In both cases we see the scan prefers $M_2$ to be around the weak scale and an order of magnitude larger than $M_1$.~Along with $\lambda_2 \sim 0.04$, and applying the supersymmetric constraints in~\eref{gmcon} and~\eref{holo1}, this implies the electroweak doublet and triplet sectors in the Higgs potential in~\eref{VGM} are largely decoupled from one another.~We see also see the fit prefers a small electroweak triplet VEV around $v_\Delta \sim 5$\,GeV for the SGM model while for the CGM model the triplet VEV is preferred to be $v_\Delta \sim 10$\,GeV.

As discussed above, in the SGM model the $M_1$ and $M_2$ mass parameters are related to the superpotential as in~\eref{muterms}.~In~\fref{muterms} we show the scan in the $(\mu$\,vs.\,$\mu_\Delta)$ plane where we see the scan prefers $\mu \sim -100$\,GeV and  $\mu_\Delta \sim 200$\,GeV.~Note that $\mu_\Delta$ largely determines the mass of the custodial fiveplet higgsino (see~\eref{Mf5}) implying a doubly charged fermion around the weak scale.~Below we show projections of these parameter scans onto the physical masses and mixing angles in the SGM and CGM models. 
\begin{figure}
\centering
\includegraphics[width=0.5\textwidth]{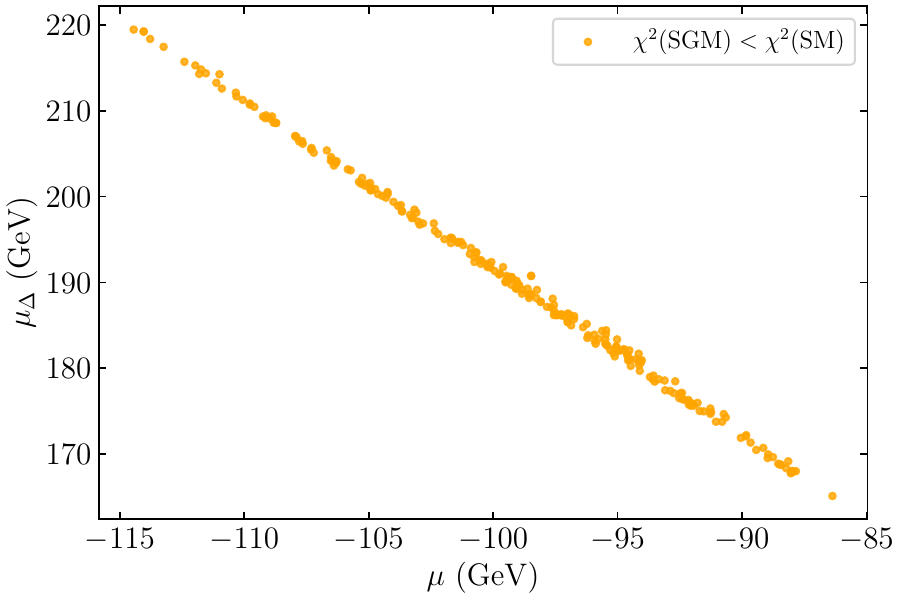}
\caption{Results of the scan for the $\mu$ terms of the superpotential in the SGM model, related to the $M_1$ and $M_2$ mass parameters as in~\eref{muterms}, in the $(\mu_\Delta$\,vs.\,$\mu)$ plane.}
\label{fig:muterms}
\end{figure}
%%%%%

%%%%%%%%%%%
%%%%%%%%%%%
%%%%%%%%%%%
\subsection{The 95\,GeV diphoton excess and 125\,GeV SM-like Higgs (custodial singlets)}

One of the results of our scan over the Higgs potential parameters is that the if we take the 95\,GeV excess to be due to a new resonance in the SGM model, it must be one of the custodial singlet scalars.~Using the same convention as in~\cite{Chen:2023bqr} we associate $h$ with the observed 125\,GeV Higgs boson (see~\eref{Higgsmix}) and $H$ with lighter custodial singlet at $95$\,GeV.~We can write the normalized couplings in terms of the Higgs and VEV mixing angles defined in~\eref{VEVmixing} and~\eref{hHmixing} respectively,
\begin{equation}
\label{eq:kappa}
\begin{aligned}
&\kappa_{V}^h = c_\alpha c_H - \sqrt{\frac{8}{3}}s_\alpha s_H,~&&
\kappa_{V}^H = \sqrt{\frac{8}{3}}c_\alpha s_H+s_\alpha c_H,\\    
&\kappa_{f}^h = \frac{c_\alpha}{c_H},~&&
\kappa_{f}^H = \frac{s_\alpha}{c_H},\\
\end{aligned}
\end{equation}
where $s_\alpha,\,c_\alpha$ corresponds to the Higgs custodial singlet mixing angle defined in~\eref{hHmixing}.~Note in the limit $\alpha \to 0$ we have $\kappa_{V}^h = \kappa_{f}^h  = 1$ for the SM-like Higgs boson at 125\,GeV (and $\kappa_{V}^H = \kappa_{f}^H  = 0$).~The loop induced normalized effective couplings to gluons and photons can be obtained in a similar way utlizing the \texttt{SPheno} code to perform the loop calculations.

%%%%%
\begin{figure}
\begin{center}
\includegraphics[scale=.49]{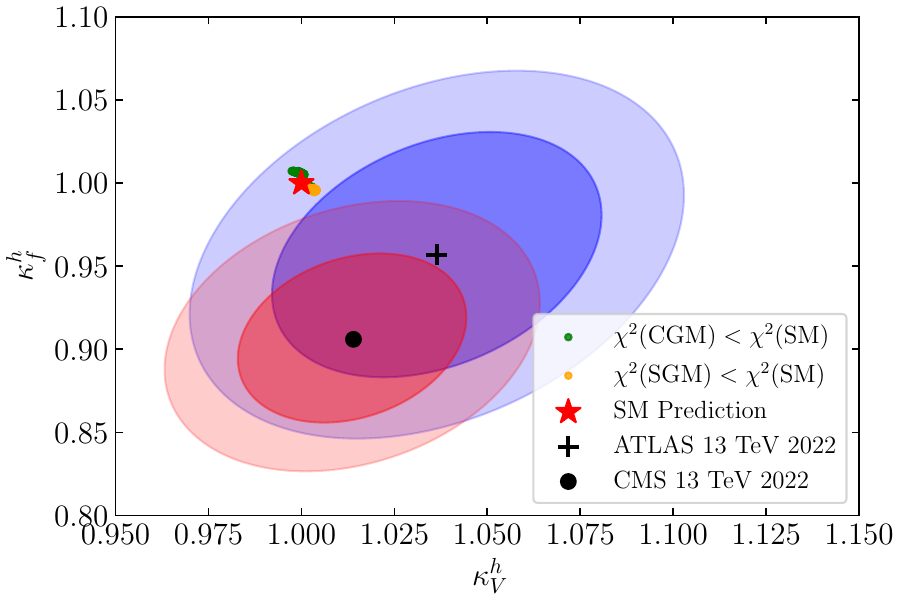}
\includegraphics[scale=.49]{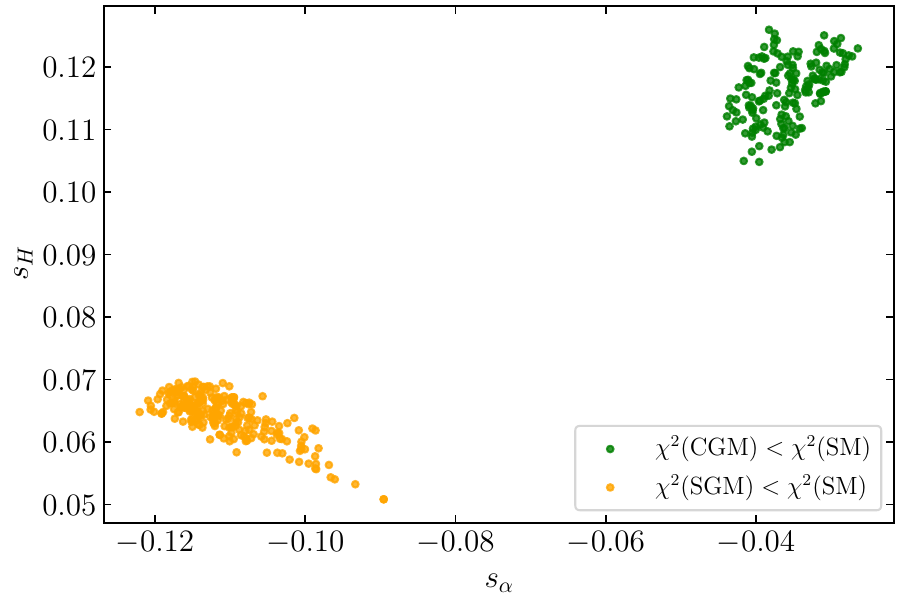}
\end{center}
\caption{{\bf Left:}\,Constraints on $(\kappa_V^h$\,vs.\,$\kappa_f^h)$ coming from measurements of the 125\,GeV Higgs boson at ATLAS (blue) and CMS (red).~We also show the allowed parameter (as consistent with data as the SM) space in the SGM (yellow) and CGM (green) models.~{\bf Right:}\,Allowed parameter space for the SGM (yellow) and CGM (green) models in the electroweak triplet VEV ($s_H$) versus custodial singlet Higgs mixing angle ($s_\alpha$) plane with the mixing angles defined in~\eref{hHmixing}.}
\label{fig:kap125}
\end{figure}
%%%%%
We first consider in~\fref{kap125} (left) constraints on $\kappa_V^h$\,vs.\,$\kappa_f^h$ coming from measurements of the 125\,GeV Higgs boson couplings~\cite{CMS:2014nkk,Khachatryan:2016vau,Sirunyan:2017tqd,Sirunyan:2017exp} at ATLAS (blue) and CMS (red).~We also show the best fit points in both the SGM (yellow) and CGM (green) models.~As we can see, measurements of the 125\,GeV Higgs boson couplings require $h$ to be SM-like (mostly electroweak doublet).~As we show on the right in~\fref{kap125}, in the SGM model this implies small VEV and Higgs mixing angles around $s_H \sim 0.06$ and $s_\alpha \sim -0.11$ which leads to $\kappa_V^H \sim -0.04$ and $\kappa_f^H \sim -0.11$.~So we see the SGM model predicts $H$ gives a $\sim (5 - 7)\%$ contribution to EWSB with negative couplings to both SM fermions and weak gauge bosons.~In the CGM model we find $s_H \sim 0.12$ and $s_\alpha \sim -0.04$ which leads to $\kappa_V^H \sim 0.15$ and $\kappa_f^H \sim -0.04$. So we see in the SGM model the coupling of $H$ to $WW$ and $ZZ$ gauge bosons enters primarily via Higgs mixing versus in the CGM model where it enters mostly due to VEV mixing.~This leads to larger couplings to fermions in the SGM model than in the CGM and allows for a sizable contribution from gluon fusion in the case of the SGM model (see~\fref{DY}).

To include the 95\,GeV diphoton excess we can write the signal strengths for the $95$ GeV resonance in terms of the normalized effective couplings as,   
\begin{equation}\label{eq:mu95}
\mu_{\gamma\gamma}=\mu_{\gamma\gamma}^{\rm LHC}
= 
((\kappa_f^H)^2  +  r_{\rm{DY}}+r_{V}  )
\frac{\calB_{\gamma\gamma,\rm model}}{\calB_{\gamma\gamma,\rm SM}},
\quad
 \mu_{bb}=\mu_{bb}^{\rm LEP}
= (\kappa_V^H)^2 
\frac{\calB_{bb,\textrm{model}}}{\calB_{bb,\textrm{SM}}},
\end{equation}
where we have only kept the contribution from gluon fusion to the SM production channels at the LHC and used $\kappa_g\simeq\kappa_f$ as it is mainly induced through a top-quark loop.~Given that the constraints coming from measurements of 125\,GeV Higgs couplings which force us into the small VEV and Higgs mixing angle regime, any single production of $H$ will be suppressed.~However, the custodial singlet $H$ can be produced in pairs with the custodial triplet $H_3$ via the Drell-Yan $q\bar{q} \to V \to H H_3$ production channel so it is also included via the ratio $r_{\rm{DY}}$ in Eq.~(\ref{eq:mu}), where the vector-scalar-scalar interaction is given by~\cite{Hartling:2014zca},
\begin{equation}
\kappa_{VHH_{3}}=\sqrt{6}s_{\alpha}s_{H}-c_{\alpha}c_{H}.
\end{equation}
As emphasized in~\cite{Delgado:2016arn,Vega:2018ddp}, this production channel is not suppressed even in the limit of small Higgs or VEV mixing since it is mediated by gauge interactions and does not depend on EWSB\footnote{The diphoton exclusion limits~\cite{ATLAS:2020pvn,CMS:2021kom,ATLAS:2022tnm} in the \texttt{HiggsTools} package are implemented through the Simplified Template Cross-Sections (STXS) framework~\cite{Berger:2019wnu,Berger:2922392}, including the gluon-gluon fusion, vector boson fusion, associated production of $VH$ and $Ht\bar{t}$, but \emph{not} the Drell-Yan Higgs pair production mechanism.}.~It is thus an important production channel for light Higgs bosons.

For $m_H = 95$\,GeV we show in~\fref{DY} the DY pair production cross section at a 13\,TeV LHC as a function of the custodial triplet mass $m_3$ in the no mixing limit $s_\alpha = s_H = 0$.
Taking $\sigma_{\rm ggF} \simeq 73$\,pb at 95\,GeV~\cite{Baglio:2022wzu}, we obtain the cross section ratio $r_{\rm{DY}} \simeq 0.01$ and the DY pair production channel will dominate for the 95\,GeV Higgs boson, implying $\kappa_f^H \lesssim 0.1$. Thus we see that current measurements of the 125\,GeV Higgs boson couplings already imply the dominant production channel of $H$ is the DY pair production channel.~Note, this production channel was not included in previous global fits of the GM model and the 95\,GeV excess~\cite{Chiang:2018cgb,Chen:2022zsh,Chen:2023bqr,Mondal:2025tzi}.~For the $b\bar{b}$ signal strength at LEP, the DY pair production channel is kinematically suppressed and the dominant production is through vector boson fusion so we have the normalized coupling to $W$ and $Z$ bosons $\kappa_V^h$.~The branching ratios in the SGM are computed with \texttt{SPheno} while those for a SM-like Higgs at 95 GeV are given by $\calB_{bb,\textrm{SM}} = 0.802,~\calB_{\gamma\gamma,\textrm{SM}}=1.39\cdot10^{-3}$ (extracted from Ref.~\cite{Denner:2011mq}).
%%%%%
\begin{figure}
\begin{center}
\includegraphics[scale=.7]{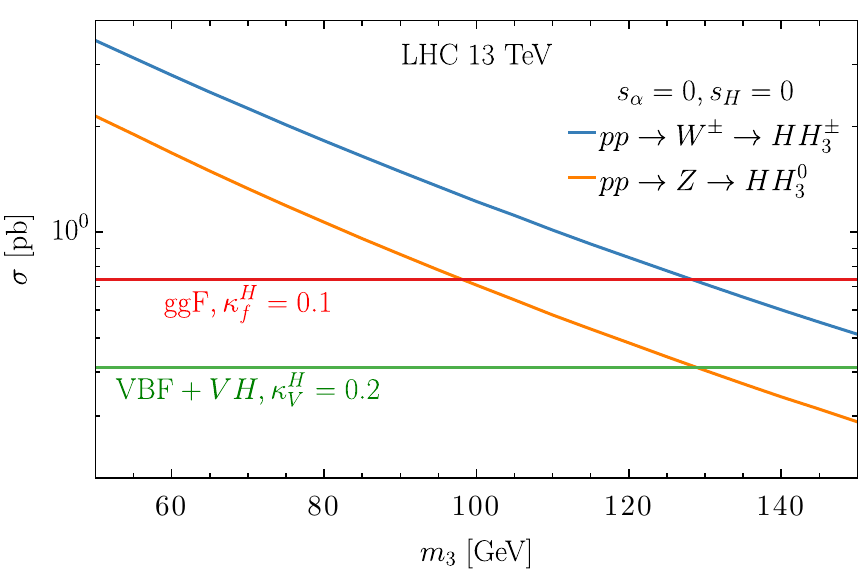}
\end{center}
\caption{Drell-Yan Higgs pair production of $H$ ($m_H = 95$\,GeV) as a function of the custodial triplet mass $m_3$ for both the $W$ (blue) and $Z$ (orange) mediated channels.~We also show the VBF + VH for $\kappa_V^H = 0.2$ (green) and gluon fusion for $\kappa_f^H = 0.1$ (red) production cross sections.}
\label{fig:DY}
\end{figure}
%%%%%. HERE
%%%%%
\begin{figure}
\begin{center}
\includegraphics[scale=.48]{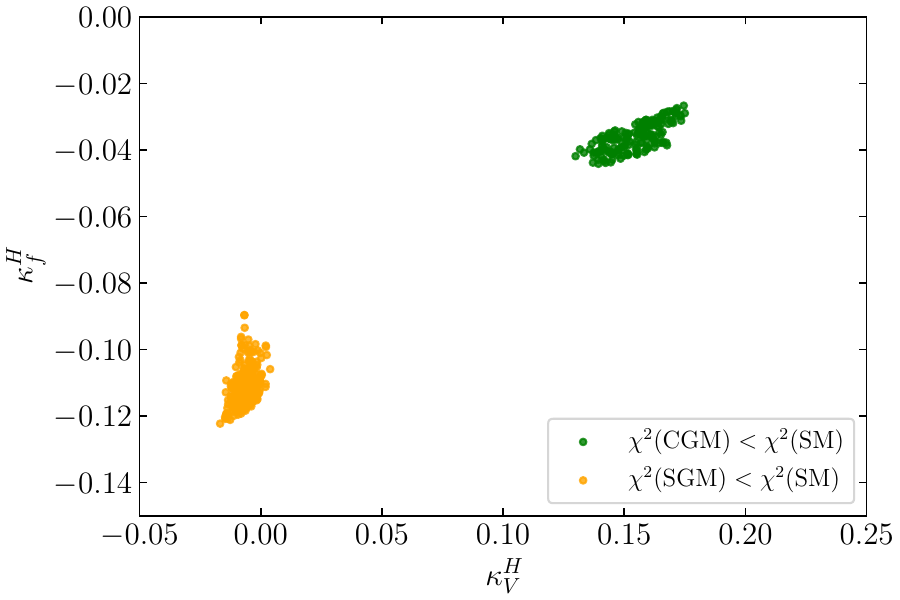}
\includegraphics[scale=.48]{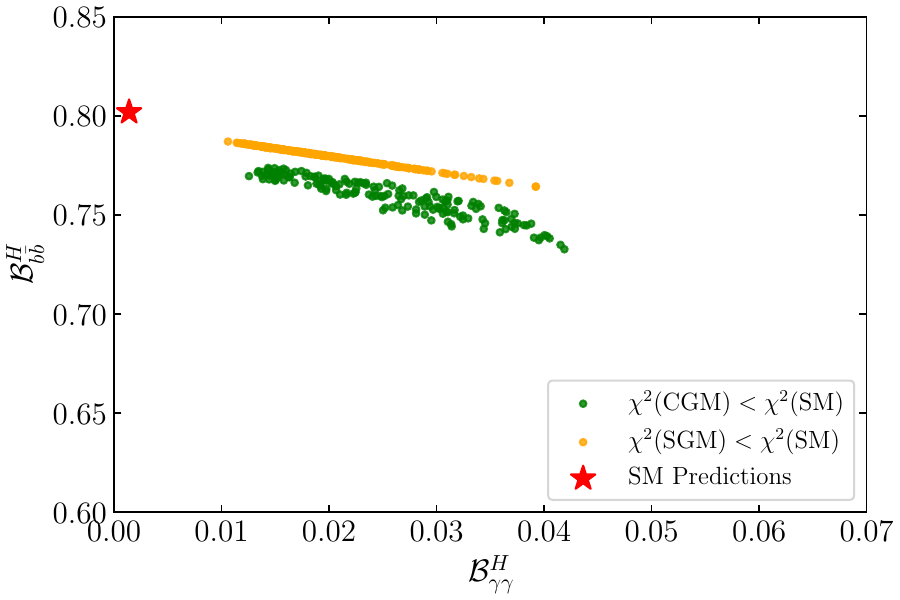}
\includegraphics[scale=0.5]{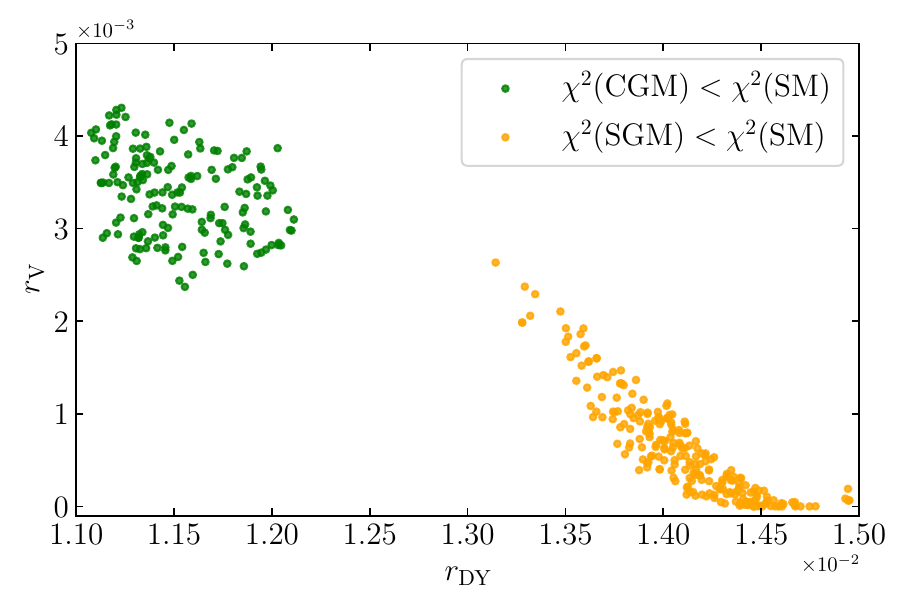}	
\end{center}
\caption{Allowed parameter space for the SGM (yellow) and CGM (green) models in the planes of $(\kappa_f^H,\kappa_V^H)$ (upper left), ($\calB_{bb}^H,\calB_{\gamma\gamma}^H$) (upper right), and $(r_{\rm DY},r_V)$  (lower).}
\label{fig:kap95}
\end{figure}

With these definitions for the signal strengths we show in~\fref{kap95} the allowed parameter space for the normalized effective couplings (left) and branching ratios (right) into $\gamma\gamma,\,b\bar{b}$ in the SGM (yellow) and CGM (green) models.~We see the SGM model predicts the branching ratio into bottom quarks to be $\calB_{bb} \sim 0.8$ which is close to the SM prediction at 95 GeV while into photons it predicts $\calB_{\gamma\gamma} \sim 0.05$ which is about 50 times larger than the SM, thus able to to explain the diphoton excess.~An enhanced effective coupling to photons is possible in the SGM and GM models because of the loop contribution from the doubly charged scalar in the custodial fiveplet Higgs which interferes \emph{constructively} with the $W^\pm$ boson contribution and requires the fiveplet to be relatively light as we'll see below.

We also see that having $\calB_{bb} \sim 0.8$ requires $\kappa_V^H \sim 0.3$ in order to explain the LEP excess~\cite{LEPWorkingGroupforHiggsbosonsearches:2003ing} in $b\bar{b}$ around 95\,GeV, much larger than the predicted value of $\kappa_V^H \approx -0.04$ in the SGM model.~Thus we see it is difficult to explain the LEP $b\bar{b}$ excess with the custodial singlet in the SGM model.~We also expect to be the case for other extended Higgs sectors invoked to explain the 95\,GeV excess though further study is needed.~As we can also see, the CGM model has a similar, but slightly larger parameter space which is consistent with the data than the SGM model preferring a slightly smaller branching ratio into $b\bar{b}$.~Note that it is difficult to obtain the needed enhanced coupling to photons in extended Higgs sectors with only electroweak doublets since they do not contain a doubly charged scalar.

%%%%%%%%%%%
%%%%%%%%%%%
%%%%%%%%%%%
\subsection{The custodial triplet and fiveplet Higgs bosons}

The phenomenologically most striking feature of the GM model is the presence of the custodial fiveplet Higgs boson which contains a neutral, charged, and doubly charged component.~Since it does not couple to fermions, the dominant production channel is DY Higgs pair production ($q\bar{q} \to V^\ast \to H_5 H_5$).~Combining DY pair production with inclusive diphoton searches at the LHC~\cite{ATLAS:2014jdv,ATLAS:2021uiz}, strong bounds can be placed on the fiveplet branching ratio into photons~\cite{Delgado:2016arn,Vega:2018ddp} versus mass which we show in~\fref{h5gaga} (left).
%%%%%
\begin{figure}[tbh]
\begin{center}
\includegraphics[scale=.52]{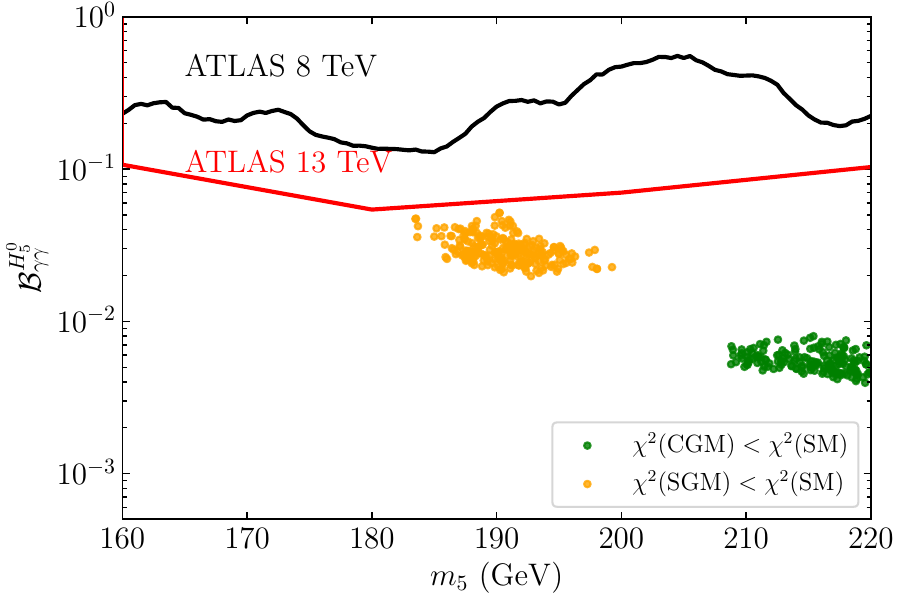}
\includegraphics[scale=.52]{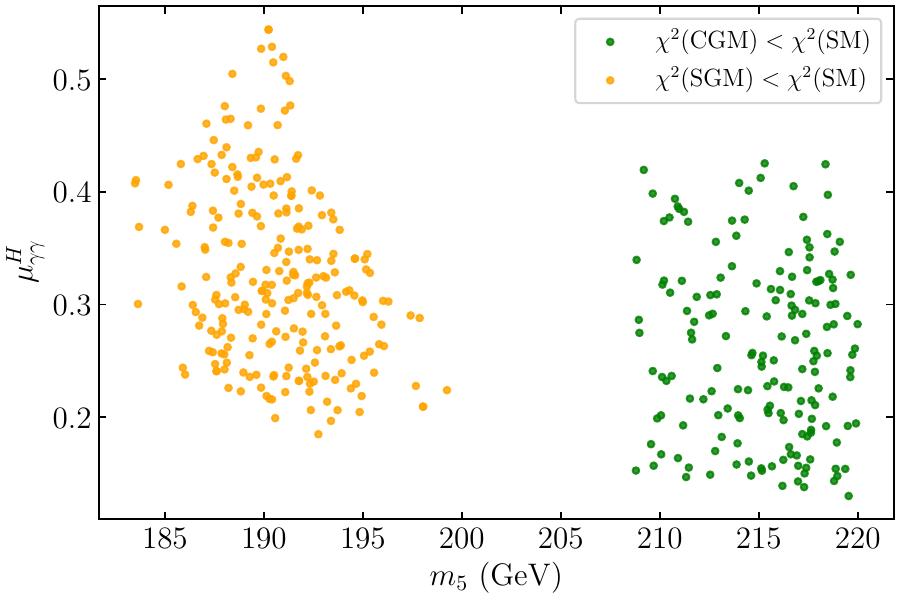}
\end{center}
\caption{{\bf Left:\,}Allowed parameter space (as consistent with data as the SM) of the custodial fiveplet branching ratio into diphotons versus $m_5$ for the SGM (yellow) and CGM (green) models.~We also show the bound coming from ATLAS 8\,TeV (black)~\cite{ATLAS:2014jdv} and 13\,TeV (red)~\cite{ATLAS:2021uiz} diphoton searches.~{\bf Right:\,}Allowed parameter space in the diphoton signal strength versus $m_5$ plane.}
\label{fig:h5gaga}
\end{figure}
%%%%%
We see the ATLAS 8 TeV~\cite{ATLAS:2014jdv} and 13 TeV~\cite{ATLAS:2021uiz}\footnote{The previous ATLAS 13 TeV release~\cite{ATLAS:2017ayi} with a smaller integrated luminosity explored in~\cite{Ismail:2020zoz,Ismail:2020kqz} only applies to the diphoton mass window $m_{\gamma\gamma}\in[200,2700]$~\GeV.} diphoton searches put a bound of $\sim 10\%$ on the fiveplet diphoton branching ratio into photons in this mass range.~As emphasized in~\cite{Delgado:2016arn,Vega:2018ddp}, since the fiveplet is dominantly produced via DY Higgs pair production which depends only on gauge interactions, these are robust bounds which are independent of the electroweak triplet VEV (or Higgs mixing angles).~The SGM model predicts the fiveplet branching ratio to be $(3 - 5)\%$ in the mass range $\sim(185 - 195)$\,GeV.~In the CGM model the branching ratio is predicted to be below $\sim 1\%$ between $\sim(210 - 220)$\,GeV.~With the current LHC 13.6 TeV~\cite{Knolle:2023vbw,Hugli:2024ofb} and future HL-LHC~\cite{Cepeda:2019klc} runs, the remaining SGM model parameter space should be directly probed and possibly entirely excluded in combination with Higgs precision measurement at future $e^+e^-$ colliders,\,including ILC~\cite{Bambade:2019fyw},\,CEPC~\cite{CEPCStudyGroup:2018ghi},\,and FCC-ee~\cite{FCC:2018evy}.

On the right in~\fref{h5gaga} we show the diphoton signal strength for $H$ at $95$\,GeV as a function of the fiveplet mass.~Since the fiveplet contains the doubly charged Higgs scalar, it is primarily responsible for the enhanced coupling to photons.~We see the SGM model prefers a somewhat lighter fiveplet Higgs around $\sim 190$\,GeV versus the CGM model for the which the fiveplet is $\sim 215$\,GeV.~Here we see the influence of the (doubly charged) fiveplet Higgsinos which interfere destructively with the fiveplet scalar, thus requiring lower $m_5$ to obtain the same enhancement as in the CGM model where there are no Higgsinos. 

In~\fref{m3m5} (left) we present the allowed parameter space in the ($m_5$\,vs.\,$m_3$) plane again for the SGM (yellow) and CGM (green) models.~As in the GM model~\cite{Chen:2023bqr}, we see in the SGM model the approximate decoupling mass relation when $s_H,s_\alpha \approx 0$,
\begin{equation}
2m_H^2\approx 3m_{H_3}^2-2m_{H_5}^2.
\end{equation}
We also see in the SGM model, explaining the 95\,GeV diphoton excess with a custodial singlet implies a custodial fiveplet, including the doubly charged component, approximately in the mass range $(185 - 195)$\,GeV while in the CGM model the range is $(210 - 220)$\,GeV.~The custodial triplet is constrained in the SGM model to be between about $(133 - 140)$\,GeV while in the CGM model the range is $(145 - 150)$\,GeV.
%%%%%
\begin{figure}[tbh]
\begin{center}
\includegraphics[scale=.52]{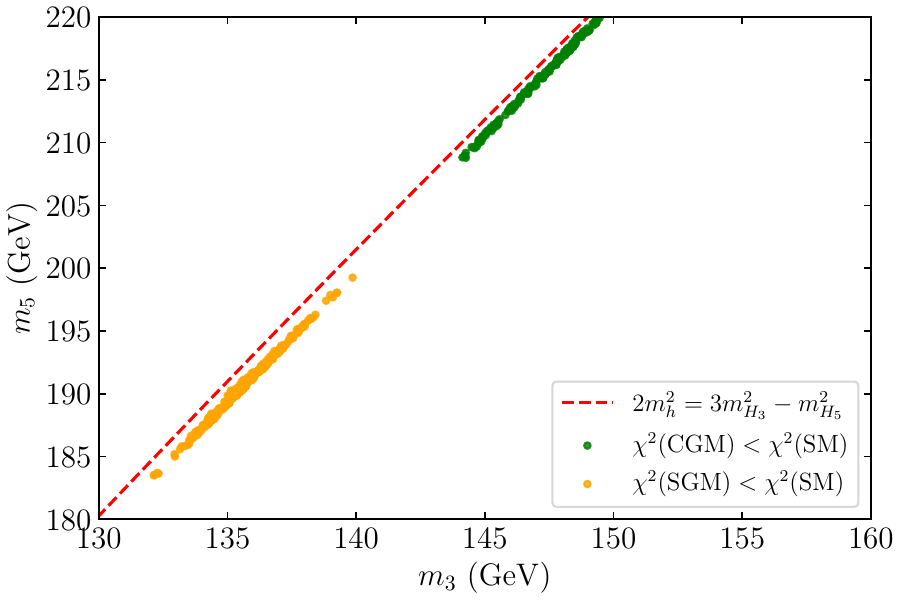}
\includegraphics[scale=.52]{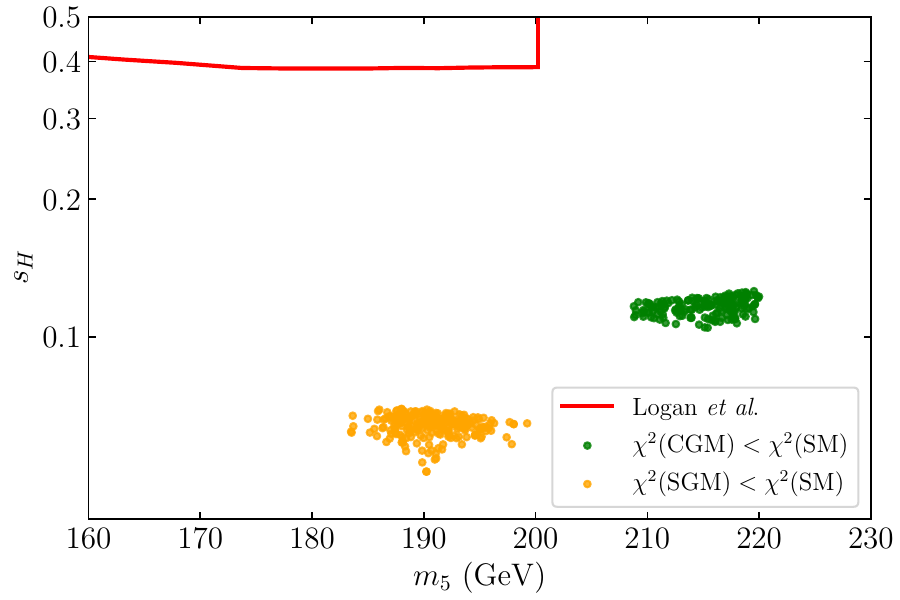}
\end{center}
\caption{{\bf Left:}\,Allowed parameter space for the SGM (yellow) and CGM (green) models in the custodial fiveplet versus triplet mass plane.\,{\bf Right:}\,Allowed parameter space in the $(s_H$\,vs.\,$m_5)$ plane where the red line indicates the bound obtained in~\cite{Ismail:2020kqz,Ismail:2020zoz} for the general GM model.}
\label{fig:m3m5}
\end{figure}
%%%%%

There are also strong bounds on the fiveplet mass coming from decays of the doubly charged component into like sign $W$ bosons, but which depend on the triplet VEV~\cite{Hartling:2014zca}.~On the right in~\fref{m3m5} we show the allowed parameter space in the $(s_H$\,vs.\,$m_5)$ plane where the most stringent constraint comes from searches at ATLAS for same-sign of $W$ boson production~\cite{ATLAS:2014jzl}.~We again see the lower preferred mass in the SGM compared to the CGM model due to the presence of the Higgsino fermions.~We also show the bound obtained in~\cite{Ismail:2020kqz,Ismail:2020zoz} for the general GM model (red line) where the supersymmetric constraint in~\eref{gmcon} is \emph{not} imposed.~We see clearly in the CGM model (green) how the supersymmetric condition on the Higgs potential greatly reduces the parameter space as well as how the presence of the custodial higgsinos in the SGM model (yellow) pushes us to lower fiveplet masses and smaller electroweak triplet VEV.

%%%%%%%%%%%
%%%%%%%%%%%
%%%%%%%%%%%
\subsection{The custodial Higgsinos}

In general we have seen that, when trying to accommodate the 95\,GeV excess, the fermion Higgsinos in the SGM model lower the mass scale of the custodial Higgs bosons as compared to the CGM model and in particular of the fiveplet Higgs boson which contains the doubly charged scalar.~This is because the Higgsinos enter through loops in the diphoton effective coupling which, as discussed above, requires enhanced couplings to photons.~Specifically, the doubly charged Higgsino fermions interfere destructively with the $W$ boson and doubly charged scalar loops which generate the effective coupling of $H$ to photons.~Thus, the doubly charged scalar in the SGM model must be somewhat lighter (as seen in~\fref{h5gaga}) to compensate for the destructive interference from the doubly charged Higgsino.

As discussed in~\sref{fspec}, in the limit of small electroweak triplet VEV to which the data forces us, the custodial fiveplet Higgsino mass is essentially a direct measure of the $\mu_\Delta$ term of the SCTM superpotential~\cite{Vega:2017gkk} which is related to $M_2$ in the GM Higgs potential (see~\eref{VGM}) via~\eref{muterms}.~We show on the left in~\fref{higgsino} the mass of doubly charged component from the custodial fiveplet versus the mass of the lightest supersymmetric particle (LSP).~We see the custodial fiveplet must be in the mass range $(170 - 220)$\,GeV which is $\approx \mu_{\Delta}$ (see~\fref{params}) giving a direct probe of the SUSY scale for the electroweak triplet sector.~We find in our scans that the LSP corresponds to the neutral component of the custodial triplet $f_3$ in~\eref{Mf3} and is preferred to be in the range $(117 - 135)$\,GeV.~It has a very small splitting with the charged component, due to hypercharge interactions, which we show in~\fref{higgsino} (right) as a function of the LSP mass.~This small mass difference suppresses the charged fermion decay through the 3-body off-shell $W$ channel $f_3^+\to f_3^0(W^{+*}\to jj/\ell\nu)$ down to $\Gamma\sim10^{-10}~\GeV(c\tau\sim\mu\textrm{m})$, which is still in the short-lived range and evades LHC searches for heavy stable charged particles~\cite{CMS:2016kce,ATLAS:2019gqq,ATLAS:2022pib,CMS:2024nhn}.~There is also the presence of a custodial singlet higgsino which is found in our scans to be the NLSP and about $\sim 20$\,GeV heavier than the custodial triplet LSP.~Note however, if one is not interested in explaining the 95\,GeV excesses, then the higgsinos can be much lighter, in principle all the way down to the direct search bounds on charged leptons of $\sim100$\,GeV or even as low as $\sim 75$\,GeV~\cite{Egana-Ugrinovic:2018roi}.~In general, due to their compressed mass spectra~\cite{Schwaller:2013baa,Ismail:2016zby}, the soft decay products of the custodial fermions makes their detection challenging and their search would benefit from dedicated strategies for compressed mass spectra~\cite{ATLAS:2017vat,ATLAS:2019lng,CMS:2019zmn,CMS:2024wzb} which we defer for future work.
 %%%%%
\begin{figure}[tbh]
\begin{center}
\includegraphics[scale=.49]{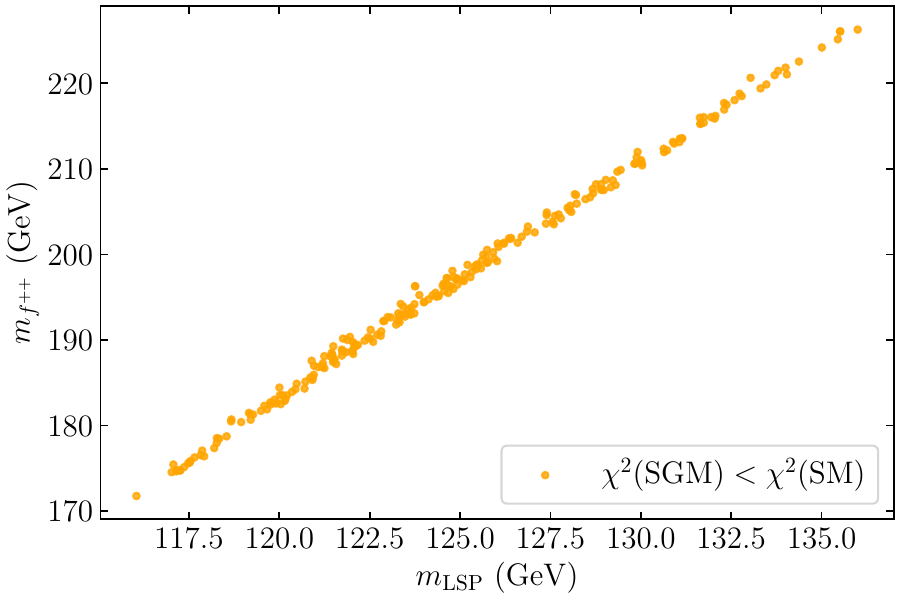}
\includegraphics[scale=.49]{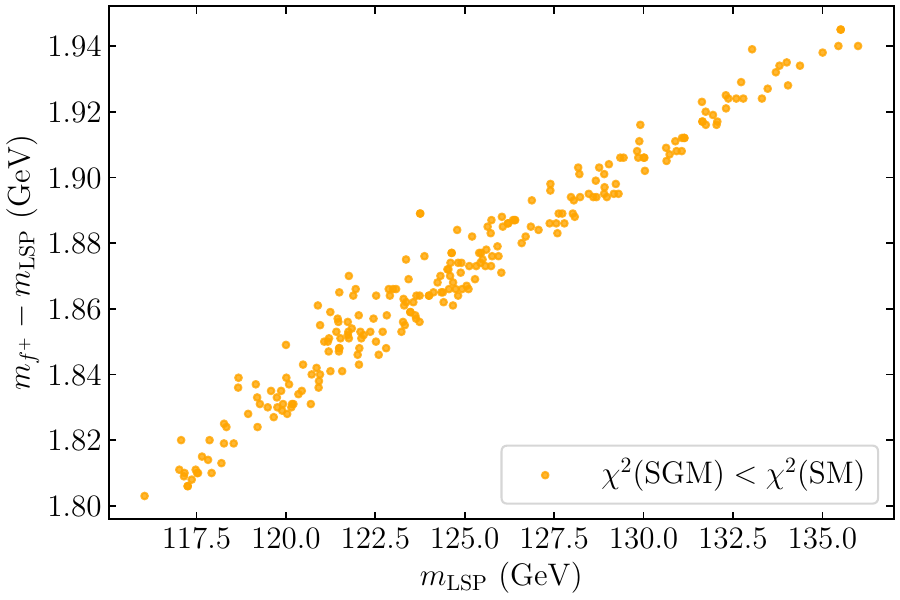}
\end{center}
\caption{{\bf Left: }Allowed parameter space in the SGM model in the doubly charged higgsino (custodial fiveplet) mass versus LSP (neutral component of the lightest higgsino custodial triplet) mass plane.\,{\bf Right:} Allowed parameter space of mass splittings between the neutral (LSP) and charged components of the higgsino custodial triplet.}
\label{fig:higgsino}
\end{figure}
%%%%%

%%%%%%%%%%%%%%%
%%%%%%%%%%%%%%%
%%%%%%%%%%%%%%%

\section{Conclusion}
\label{sec:conclusion}

In this study we have examined whether the recently reported excesses around 95\,GeV by ATLAS~\cite{ATLAS:2018xad} and CMS~\cite{CMS:2017yta,CMS:2023yay} in searches for di-photon resonances could be the first hints of an extended electroweak symmetry breaking (EWSB) sector in the supersymmetric Georgi-Machacek model (SGM)~\cite{Vega:2017gkk}.~To assess this we have performed a global fit of the SGM model to all relevant data including measurements of the observed 125 GeV Higgs boson and found the regions of parameter space (see~\fref{params}) where the SGM can fit the current data at least as well as the Standard Model (SM).~We find that the SGM model can acomodate the data if the 95\,GeV diphoton excesses are due to the lightest custodial singlet Higgs boson (mostly electroweak triplet) while the heavier singlet is the observed (mostly SM-like electroweak doublet) 125\,GeV Higgs boson (see~\fref{kap125}).~Relative to a SM Higgs boson at the same mass, the 95\,GeV Higgs must have enhanced couplings to photons and suppressed couplings to SM fermions to explain the excesses while its vacuum expectation value is predicted to contribute $\sim(5 - 7)\%$ to EWSB.~We have included in our analysis the Drell-Yan Higgs pair production channel, which is not present in the SM and was neglected in previous studies, and found it to be the dominant production channel (see~\fref{DY}).~We have also included $t\bar{t}H_3^\pm \to \tau^\pm \nu$ searches in our analysis which were also not included in previous fits to the 95\,GeV excesses.~Including both the Drell-Yan pair production channel and $t\bar{t}H_3^\pm \to \tau^\pm \nu$ searches severely constrains the SGM model parameter space which is expected to also be the case for any other extended EWSB sector.

Since supersymmetry highly constrains the SGM model Higgs potential, the rest of the scalar (Higgs boson) and fermion (higgsino) spectrum is predicted within narrow mass windows allowing for targeted searches.~If the 95\,GeV diphoton excesses are due to the lighter custodial singlet, then the SGM predicts a doubly charged scalar in the mass range $\sim(185 - 195)$\,GeV (see~\fref{h5gaga}) as well as a doubly charged fermion in the range $\sim(170 - 220)$\,GeV (see~\fref{higgsino}) which are part of the custodial fiveplet Higgs and Higgsino respectively.~The CP odd custodial triplet Higgs (scalar) is predicted to be in the range $\sim(133 - 140)$\,GeV  (see~\fref{h5gaga}).~The global fit also points to a fermion LSP which is the neutral component of the custodial triplet Higgsino with a mass in the range $\sim(117-135)$\,GeV and a very small splitting with the charged component (see~\fref{higgsino}) while the custodial singlet neutral Higgsino is predicted to be around $\sim 20$\,GeV heavier.~The mass of the custodial fiveplet higgsino is a direct probe of the supersymmetry scale of the electroweak triplet sector and points to $\mu_\Delta \sim 200$\,GeV.

As part of our analysis we also compared the parameter space in the SGM model with a constrained version of the non-supersymmetric GM model (dubbed the constrained GM (CGM) model) given by the conventional GM model, but with the supersymmetry constraints in~\eref{gmcon}\,-\,\eref{holo2} applied to the GM Higgs potential in~\eref{VGM}.~This allowed us to isolate the effects of the Higgsino fermions in the SGM model and see to what extent they affect the parameter space of the global fits.~We found that the presence of the higgsinos lowers the mass scale of the Higgs boson sector and in particular the fiveplet Higgs boson.~This is because the doubly charged higgsino suppresses the loop induced coupling of the 95\,GeV Higgs boson to photons due to destructive interference with the $W^\pm$ boson and doubly charged scalar (custodial fiveplet) loops.~In order to accommodate the diphoton excesses, this pushes the custodial fiveplet Higgs to be lighter in the SGM model ($\sim 190$\,GeV) as compared to the CGM model ($\sim 215$\,GeV) (see~\fref{h5gaga}).~Finally, the contribution to EWSB can be slightly larger in the CGM model ($\sim 10\%$) versus the SGM model ($\sim 5\%$).

In summary, if the SGM model is responsible for the 95\,GeV diphoton excesses, we have a tightly constrained and predictive parameter space.~This would point to a variety of new Higgs bosons below the weak scale with tightly correlated mass spectra and decay patterns.~It would also point to a custodial higgsino sector at around the same mass scale as the Higgs bosons, as their masses are correlated through the Higgs potential, with nearly degenerate mass spectra.~Should the excesses persist, dedicated searches which are optimized for compressed spectra will become more motivated.~Ultimately, while correlations between the custodial Higgs bosons can give a strong indicator of its supersymmetric origins, the SGM model can only be unambiguously established if the custodial higgsinos are discovered.~We expect that current LHC search strategies designed for the GM model can be easily adopted to search for the Higgs sector in the SGM model, in addition to dedicated searches for the custodial fermions.

\begin{acknowledgements}
\noindent
The work of R.V.M. has been partially supported by Junta de Andaluc\'a Projects\,P21-00199, A-FQM-472-UGR20 (fondos FEDER) and by SRA (10.13039/501100011033) and ERDF under grant PID2022-139466NB-C21.
The work of KX is supported by the U.S. National Science Foundation under Grants No.~PHY-2310291 and PHY-2310497.
The work of KX was performed partly at the Aspen Center for Physics, which is supported by the U.S. National Science Foundation under Grant No. PHY-1607611 and PHY-2210452. We have used the High-Performance Computing resources at SMU M3 and MSU HPCC.

\end{acknowledgements}

\bibliographystyle{utphys}
\bibliography{reference}
\end{document}